\documentclass[aps,prd,twocolumn,superscriptaddress]{revtex4}
\usepackage{graphicx}
\usepackage{amsmath}
\begin{document}

\title{Gravitational resonance spectroscopy with an oscillating magnetic field gradient in the GRANIT flow through arrangement.}

\def\LPSC{LPSC, Universit\'e Grenoble-Alpes, CNRS/IN2P3, Grenoble, France}
\def\UV{Physics Department, University of Virginia 382 McCormick Road, Charlottesville, VA 22904, U.S.A.}
\def\ILL{Institut Max von Laue - Paul Langevin, 71 av. des Martyrs, 38000 Grenoble, France}
\def\LI{Lebedev Institute 53 Leninskii pr., 119991 Moscow, Russia}

\author{G. Pignol}
\email{guillaume.pignol@lpsc.in2p3.fr}
\affiliation{\LPSC}

\author{S. Bae{\ss}ler} \affiliation{\UV}
\author{V.~V.~Nesvizhevsky} \affiliation{\ILL}
\author{K. Protasov} \affiliation{\LPSC}
\author{D. Rebreyend} \affiliation{\LPSC}
\author{A.~Yu.~Voronin} \affiliation{\LI}

\date{\today}

\begin{abstract}
Gravitational resonance spectroscopy consists in measuring the energy spectrum of bouncing ultracold neutrons above a mirror by inducing resonant transitions between different discrete quantum levels. 
We discuss how to induce the resonances with a flow through arrangement in the GRANIT spectrometer, excited by an oscillating magnetic field gradient. 
The spectroscopy could be realized in two distinct modes  (so called DC and AC) using the same device to produce the magnetic excitation. 
We present calculations demonstrating the feasibility of the newly proposed AC mode. 
\end{abstract}

\maketitle

\section{Introduction}

Ultracold neutrons bouncing over a horizontal mirror are used to probe quantum effects of a particle in the gravitational field \cite{ReviewQuantumStates}. 
The vertical motion of such neutrons bouncing at sub-millimeter distance from the mirror has discrete energy spectrum 
that can be calculated from the stationary Schr\"odinger equation
\begin{equation}
\label{stationary}
\frac{\hbar^2}{2 m} \frac{d^2}{dz^2} \psi_n + mg z \, \psi_n = E_n \psi_n, 
\end{equation}
where $m$ is the neutron mass, $g = 9.81$~m/s$^2$ is the local gravitational acceleration, $\psi_n(z)_{n=1,2,\cdots}$ are the stationary wavefunctions with associated energy $E_n$. 
The existence of the quantization of the vertical motion was demonstrated a decade ago \cite{nature}, 
profiting from the relatively large spatial extension of the ground state wavefunction characterized by $z_0 = (\hbar^2/2m^2g)^{1/3} \approx 5.87 \ \mu$m. 

Precision study of the quantum states is motivated by their sensitivity to extra short range interactions (see \cite{antoniadis} and references therein) in particular those induced by Chameleon Dark Energy \cite{chameleons1,chameleons2,chameleons3}. 
In addition, the neutron quantum states provide a unique test of the weak equivalence principle in a quantum regime, since the inertial and gravitational masses in eq. (\ref{stationary}) do not cancel. 

High precision measurements can be achieved with the gravitational resonance spectroscopy technique, where transitions between quantum states are induced by a periodic excitation \cite{QBounce}. 
The characteristic frequencies of the transitions, as low as $f_0 = m g z_0 / 2 \pi \hbar \approx$~145~Hz, are accessible by electrical as well as mechanical oscillators. 
More precisely, the quantum frequencies for a transition $n \rightarrow m$ from the state of energy $E_n$ to the state of energy  $E_m$ is given by $f_{nm} = (E_n - E_m) / 2 \pi \hbar$. 
Solving the problem (\ref{stationary}) one can show that
\begin{equation}
\label{trueFrequencies}
f_{nm} = f_0 (\epsilon_n - \epsilon_m), 
\end{equation}
where $\epsilon_n = \left[ 2.338, 4.088, 5.521, 6.787, \cdots \right]$ is the series of the negative zeros of the Airy function. 
To perform the spectroscopy of the bouncing neutron, the interaction that couples different quantum states can be a vibration of the bottom mirror, or an oscillating magnetic field gradient \cite{nesvizhevskyProtassov}. 
The former has been used by the QBounce collaboration \cite{QBounce}, the latter will be used in the GRANIT spectrometer \cite{GRANIT}. 

As a first step of GRANIT, a flow through measurement of the resonant transitions magnetically excited between the first three quantum states will be realized as first proposed in \cite{Kreuz}. 
According to this proposal, a space periodic (but static) magnetic field gradient will be generated at the surface of the bottom mirror. 
The frequency of the excitation seen by a neutron will thus vary according to its horizontal velocity. 
A detailed analysis of this scheme is provided in \cite{GRANIT}. 
Another mode of operation (the AC mode) can be implemented with the same setup, consisting in generating a homogeneous gradient, but oscillating in time, that should allow a more direct probe of the resonances. 

The paper is organized as follows: 
in section 2 we present the flow through setup, 
in section 3 we describe in some details the magnetic excitation, 
in section 4 we calculate the conditions for the adiabaticity of spin transport and
in section 5 we present a theoretical description of the magnetically induced transitions in the AC mode.

\section{The GRANIT flow through setup}

\begin{figure}
\includegraphics[width=0.97\linewidth]{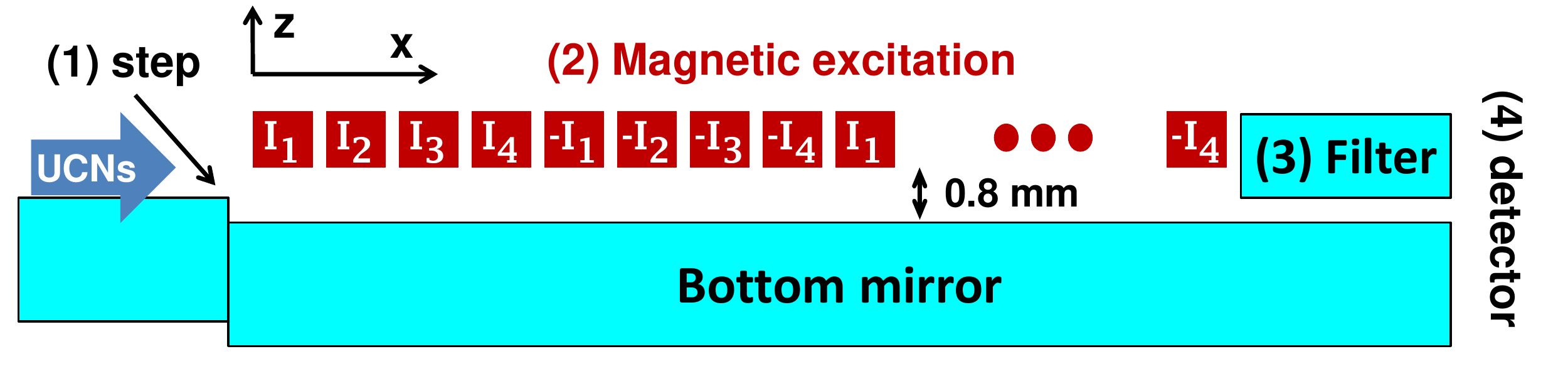}
\caption{
Sketch of the flow through setup. Ultracold neutrons enter from the left, they go through the step to depopulate the ground quantum state (1), 16~cm long transition region (2), 9~cm long analyzer (3) and detector (4). 
}
\label{sketch}
\end{figure}

The sketch of the flow through setup is shown in fig. \ref{sketch}. 
Ultracold neutrons are produced from a dedicated superthermal source installed at a cold beamline of the high flux reactor of the Institut Laue Langevin. 
The source relies on down-scattering of neutrons with a wavelength of $0.89$~nm in a superfluid helium bath cooled down to 0.8~K \cite{Schmidt-Wellenburg,Zimmer}. 
According to a preliminary measurement of the UCN velocity spectrum \cite{Roulier}, we expect the $v_x$ velocity along the beam to be distributed with a mean value of $v_0 = 4$~m/s and a standard deviation of $1.5$~m/s. 
In the remaining of this article we will assume a Gaussian profile for the $v_x$ distribution with those parameters. 
UCNs are extracted using a narrow slit to accept only those with practically no vertical velocity. 
The spectroscopy is performed with four steps: (1) state preparation, (2) resonant transition, (3) state analysis, (4) detection of transmitted flux. 

\begin{itemize}
\item
UCNs are first prepared in an excited state by going down a step (1) of height $15~\mu$m. 
The populations $p_n$ of the quantum states after the step are expected to be about $p_1 = 0.02, p_2 = p_3 = p_4 = 0.3$. 
Thus, the population of the ground quantum state is suppressed as compared to the populations of excited states. 
\item
Next, transitions between quantum states are induced with a periodic magnetic field gradient. 
The length of the transition region is $L = 16$~cm, corresponding to an average passage time $t_0 = 40$~ms. 
Two different schemes could be implemented in principle: the AC excitation and the DC excitation. 
In the DC mode, the field gradient is static and spatially oscillating in the $x$ direction with a period of $d = 1~$cm. 
In this case only neutrons with specific horizontal velocities meet the resonance condition. 
The deexcitation $2 \rightarrow 1$ is expected to be induced by an excitation frequency $f_{21} = 254$~Hz, corresponding to the resonant horizontal velocity of $v_{21} = d \ f_{21} = 2.54~$m/s, 
for the $3 \rightarrow 1$ case we expect $f_{31} = 462$~Hz and $v_{31} = 4.62$~m/s. 
In the AC mode, the field gradient is spatially uniform, oscillating in time. 
One would then find the resonances by directly scanning the excitation frequency. 
\item
A second horizontal mirror above the main mirror serves as a state analyzer. 
For a slit opening of about $25~\mu$m, only ground state neutrons are accepted, higher quantum states are rejected. 
The length of the analyzer in the $x$ direction is $9$~cm. 
\item 
Finally, neutrons are detected at the exit of the analyzer. 
In AC mode, the flux of transmitted neutrons should display a resonance pattern as a function of the excitation frequency. 
In DC mode, one has to measure the horizontal velocity of the transmitted neutrons to deduce the resonant frequency. 
This is achieved by measuring the height of the neutrons with a position sensitive detector after a free fall distance of 30~cm. 
\end{itemize}

Before developing the details, let us estimate the strength of the needed oscillating magnetic field gradient. 
The interaction of a neutron with a magnetic field $\vec{B}$ is described by the operator $- \hat{\vec{\mu}} \cdot \vec{B}$, where $\hat{\vec{\mu}}$ is the neutron magnetic moment observable acting on the internal spin degree of freedom. 
When the variation of the magnetic field is slow compared to the Larmor frequency, the spin will follow the direction of the magnetic field. 
This adiabaticity condition will be addressed in section 4. 
In this case the neutron trajectory and the spin dynamics are effectively decoupled. 
Then the motion of the neutron is determined by the potential $s \mu |\vec{B}|$ where $\mu = 60.3$~neV/T is the magnetic moment of the neutron and $s = 1$ for ``spin up'' neutrons and $s = -1$ for ``spin down'' neutrons. 
Classically, a vertical force is applied on the neutron by the field gradient $\partial_z |\vec{B}|$. 
Let us now assume a magnetic excitation of the form $|\vec{B}| = \beta z \cos(\omega t)$. 
The quantum mechanical excitation potential reads
\begin{equation}
\hat{V} = s \mu \ \beta \ \hat{z} \ \cos( \omega t). 
\end{equation}
When the excitation frequency is close to a resonance, $\omega \approx 2 \pi f_{nm}$, a Rabi oscillation between states $|n \rangle$ and $|m \rangle$ will take place at the angular frequency
\begin{equation}
\label{Rabi}
\Omega_{n m} = \frac{\mu}{\hbar} \langle n|\hat{z}|m \rangle \beta, 
\end{equation} 
where the matrix elements of $\hat{z}$ can be expressed as (see e.g. \cite{vibrations})
\begin{eqnarray}
\label{matrixElements}
\langle n|\hat{z}|m \rangle & = & \frac{2 z_0}{(\epsilon_n - \epsilon_m)^2}  \quad ( n \neq m), \\
\nonumber
\langle n|\hat{z}|n \rangle & = & \frac{2}{3} \ z_0 \ \epsilon_n. 
\end{eqnarray}

To maximize the transition probability at resonance, the excitation strength $\beta$ should verify $\Omega t_0 = \pi$, where $t_0$ is the excitation time. 
This condition can be expressed using eq. (\ref{Rabi}) and (\ref{matrixElements}) as
\begin{equation}
\beta_{\rm needed} = \frac{\pi}{2} \frac{\hbar}{\mu z_0} \left( \frac{f_{nm}}{f_0} \right)^2 \ \frac{1}{t_0}.
\end{equation}
One finds a needed field gradient of $\beta = 0.22$~T/m to induce the $2 \rightarrow 1$ transition and $\beta = 0.74$~T/m to induce the $3 \rightarrow 1$ transition.


\section{The magnetic excitation}

The magnetic field excitation will be generated by an array of $128$ copper wires with square section arranged as shown in fig. \ref{sketch}. 
In practice the system is constituted of four modules, each one holding $32$ adjacent wires. 
A wire has a section of $1$~mm$^2$ and a length of $30$~cm in the $y$ direction. 
Adjacent wires are separated by a gap of $0.25$~mm. 
Electrical connectors are arranged so that the following 8-periodic pattern current could be applied $I_1, I_2, I_3, I_4, -I_1, -I_2, -I_3, -I_4, I_1 \cdots$. 
Thus the magnetic field will be 1-cm periodic. 
The system will be placed above the horizontal mirror in the transition region as shown in fig. \ref{sketch} at a distance of $0.8$~mm from the mirror. 

\begin{figure}
\includegraphics[width=0.72\linewidth,angle=90]{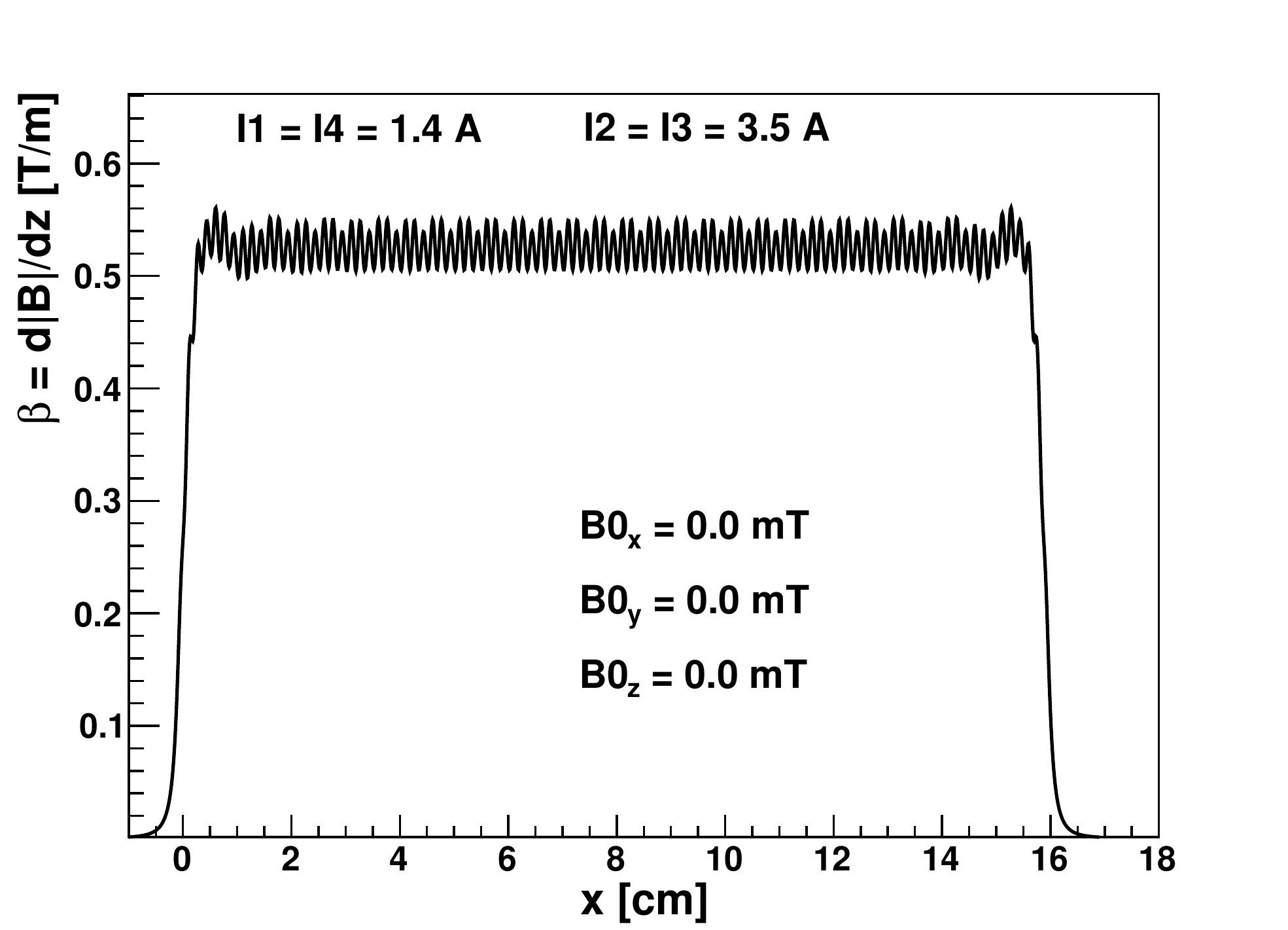}
\caption{
Magnetic field gradient $\partial_z |B|$ produced at the surface of the main mirror, without any external magnetic field. 
}
\label{gradient}
\end{figure}

The magnetic field generated by a single infinitely long square wire can be calculated analytically, the corresponding formulas are reported in the appendix. 
The magnetic field components $B_x(x), B_z(x)$ and gradients $\partial_z B_x(x), \partial_z B_z(x)$ at the surface of the mirror are obtained by summing the corresponding quantities for each of the 128 wires weighted by the individual currents. 
Then the field gradient is calculated according to
\begin{equation}
\label{grad}
\partial_z |B| = \frac{B_x \partial_z B_x + B_z \partial_z B_z}{|B|}. 
\end{equation}
It is possible to tune the currents to obtain a homogeneous gradient at the surface of the mirror. 
We show in fig. \ref{gradient} the result for $I_1 = I_4 = 1.4$~A and $I_2 = I_3 = 3.5$~A where a field gradient of $0.52$~T/m is obtained, as needed to induce resonant transitions between quantum states. 
This will be the benchmark configuration for the rest of the article. 
The residual ``noise'' seen in fig \ref{gradient} has an amplitude of $0.02$~T/m; this noise would increase for a wire array closer to the mirror. 
The frequency of this noise seen by a neutron passing at $4$~m/s is about $2$~kHz, much higher than the frequencies of interest for resonant transitions between low lying quantum states. 

\begin{figure}
\includegraphics[width=0.72\linewidth,angle=90]{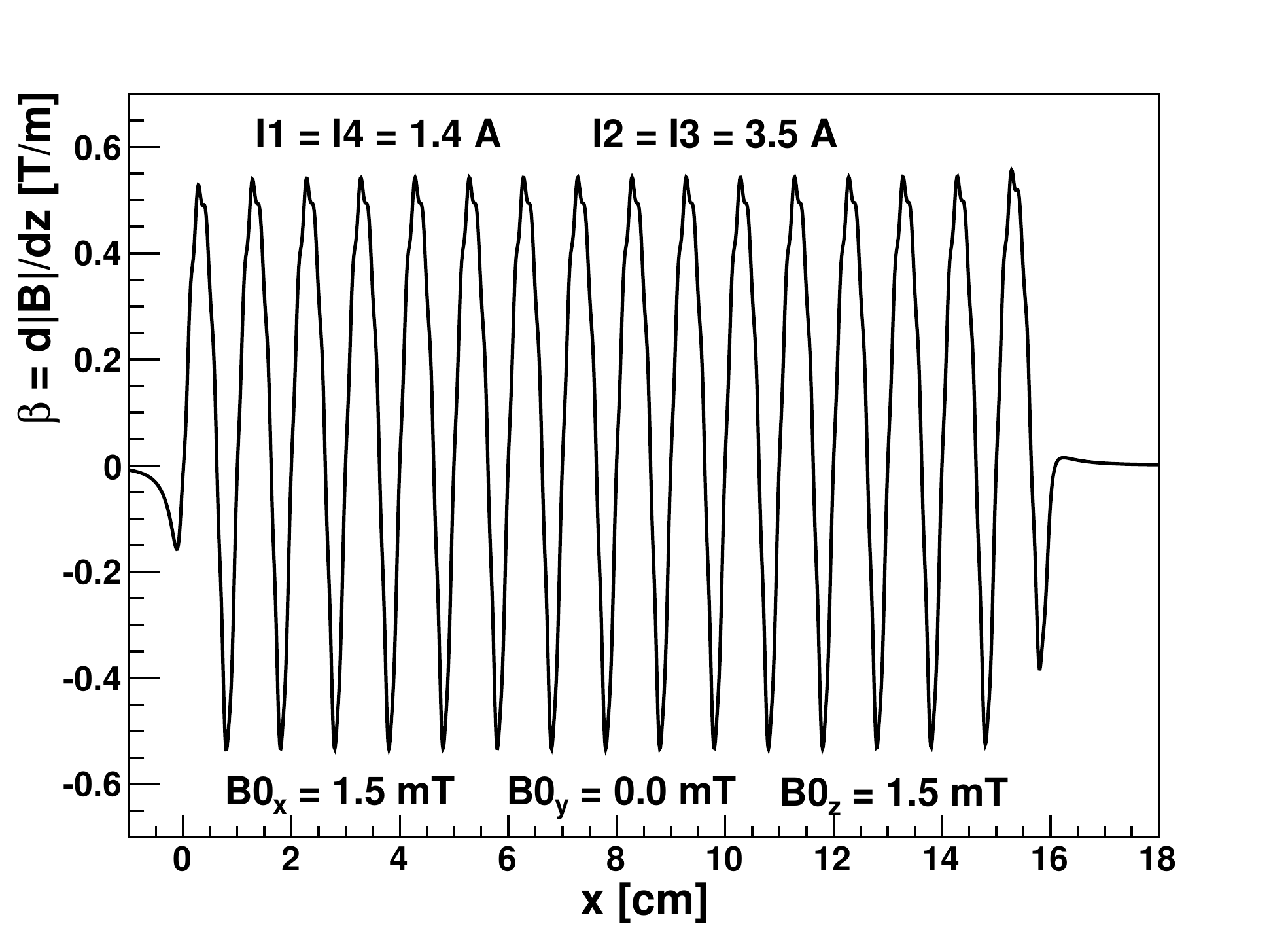}
\caption{
Magnetic field gradient $\partial_z |B|$ produced at the surface of the main mirror with external magnetic field applied in the $x,z$ plane. 
}
\label{gradient_osc}
\end{figure}

Note that the result shown in fig. \ref{gradient} assumes that no external field is applied. 
By applying an external field $\vec{B}_0$ with for example $B_{0,x} = B_{0,z} = 1.5$~mT and $B_{0,y} = 0$, the situation changes dramatically as shown in fig. \ref{gradient_osc}. 
With a strong external field applied, a gradient oscillating in the $x$ direction with a period of $1$~cm is generated. 

As a result, the wire array is a versatile device to generate the field gradient that can be used for the AC excitation mode as well as for the DC excitation mode. 
In the DC mode, we apply DC current in the wire array and apply a strong external field in the $x,z$ direction. 
The vertical force exerted by the field gradient on the neutron will oscillate in space. 
In the AC mode, we apply AC current in the wire array and a small external field $B_{0,y}$ to satisfy the adiabaticity of spin transport, as detailed in the next section. 

\begin{figure}
\includegraphics[width=0.72\linewidth,angle=90]{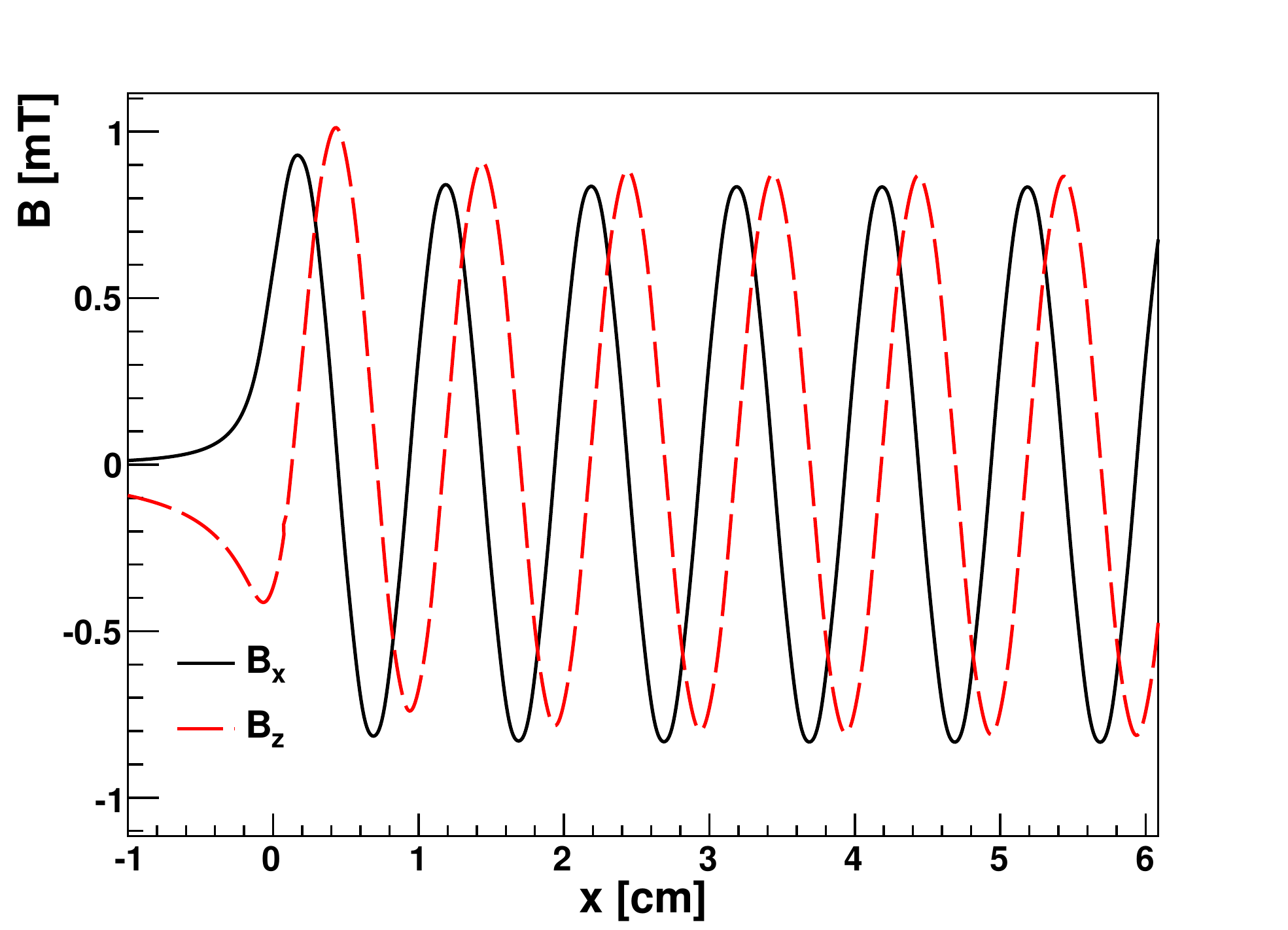}
\caption{
Magnetic field generated at the surface of the main mirror assuming the benchmark configuration $I_1=I_4 = 1.4$~A and $I_2 = I_3 = 3.5$~A.
}
\label{fieldBxBz}
\end{figure}

\section{The adiabaticity of spin transport}

The magnetic field gradient produced by the array of wires described in the previous section will exert a force on the passing neutrons. 
The sign of the force depends on the relative orientation between the neutron spin and the magnetic field. 
To induce resonant transitions between quantum states, a neutron should feel a well defined oscillating vertical force. 
Thus, one must make sure that the magnetic field is strong enough to hold the neutron spins parallel or antiparallel to the magnetic field at any time. 
If the adiabaticity condition for spin transport along the wire array is fulfilled, then the spin dynamics and the neutron trajectory are decoupled. 

In the AC mode of excitation, the input current in the wire array is oscillating with driving frequency $f$. 
The magnetic field amplitude created by the wire array at the surface of the mirror is given by the pattern shown in fig. \ref{fieldBxBz}, 
with the whole pattern oscillating in time at the driving frequency $f$. 
Thus the magnetic field generated by the wire array will not be sufficient to hold the neutron spin, since the magnitude of the field crosses zero at a frequency $f$. 
To maintain a nonzero value of the field magnitude at anytime, a static, homogeneous external field $B_{0,y}$ is applied in the transverse $y$ direction. 
The purpose of this section is to calculate the minimum $B_{0,y}$ field to apply in order to guarantee the adiabaticity of neutron spin transport when passing by the magnetic excitation. 

Here we calculate the spin dynamics only, assuming that a neutron pass below the wire array in a straight horizontal line trajectory, 
at the surface of the bottom mirror. 
A neutron with velocity $v$ along the $x$ direction sees a time-dependent magnetic field in its rest frame given by: 
\begin{eqnarray}
B_x(t) & = & B_1 \cos(2 \pi f t + \phi) \sin(2 \pi v t /d), \\
B_y(t) & = & B_{0,y}, \\
B_z(t) & = & -B_1 \cos(2 \pi f t + \phi) \cos( 2 \pi v t /d). 
\end{eqnarray}
It results from the combination of the oscillation of the field in space with period $d$ and the oscillation in time with frequency $f$. 
We will set $B_1 = 0.8$~mT for the benchmark wire currents described in the previous section. 

The spin dynamics is given by the Bloch equation for the polarization vector $\vec{\Pi}$: 
\begin{equation}
\label{Bloch}
\frac{d \vec{\Pi}}{dt} = \gamma \ \vec{\Pi} \times \vec{B}(t), 
\end{equation}
where $\gamma = 2 \mu / \hbar = 183$~kHz/mT is the neutron gyromagnetic ratio. 

\begin{figure}
\includegraphics[width=0.72\linewidth,angle=90]{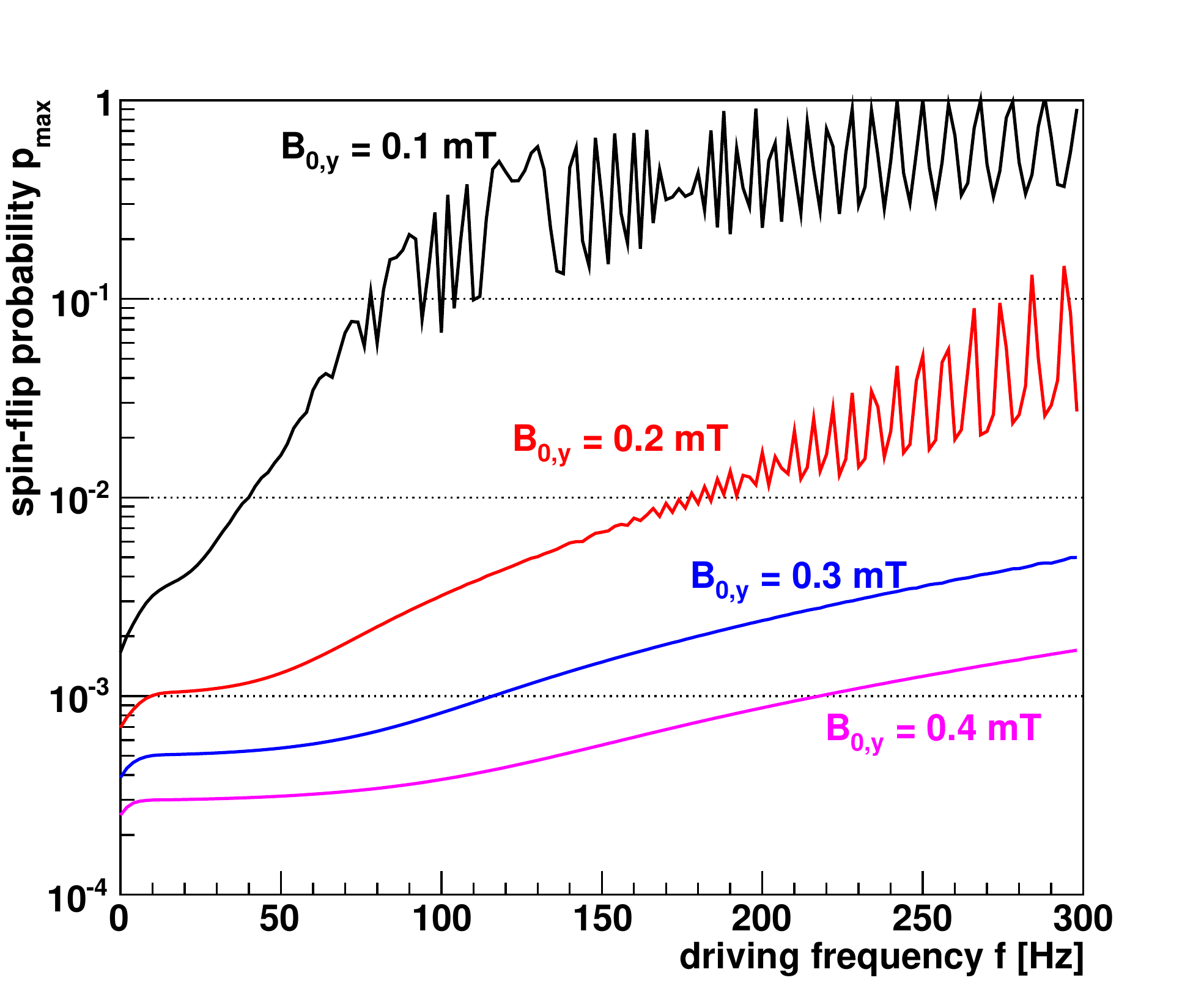}
\caption{
Numerically calculated neutron spin-flip probability during the passage below the wire array in the benchmark configuration $I_1=I_4 = 1.4$~A and $I_2 = I_3 = 3.5$~A, 
as a function of the driving frequency, for different values of the external $B_{0, y}$ field. 
}
\label{adiabaticity}
\end{figure}

We have solved numerically the Bloch equation using a Runge Kutta solver. 
The initial condition for the polarization vector $\vec{\Pi}(0)$ was set to the unit vector aligned with $\vec{B}(0)$, describing a ``spin up'' neutron. 
We define the spin-flip probability at time $t$ as $p(t) = (1 - \vec{\Pi} \cdot \vec{B} / |\vec{B}|)/2$. 
With this definition $p(t) = 0$ if the spin is aligned with the magnetic field at time $t$ and $p(t) = 1$ if the spin has reversed its direction relative to the magnetic field at time $t$. 
As an adiabaticity criterion we take $p_{\rm max}$, the maximum spin-flip probability during the passage of a neutron below the wire array of duration $L/v$. 
For a given set of parameters $B_{0,y}, f, v, \phi$ the criterion $p_{\rm max}$ was numerically calculated. 
The result was then averaged over the phase $\phi$ and the velocity spectrum $v$. 
The final result is presented in fig. \ref{adiabaticity} as a function of the driving frequency $f$, for different values of the external field $B_{0,y}$. 

As a conclusion of the numerical study, the value of the external field $B_{0, y} = 0.3$~mT 
is sufficient to hold the neutron spin with an accuracy better than one percent, 
in the frequency range of interest between 0 and $300$~Hz. 
In the following calculations the external holding transverse field will thus be set to $B_{0, y} = 0.3$~mT.

\section{The resonant transitions in the AC mode}

\begin{figure}
\includegraphics[width=0.72\linewidth,angle=90]{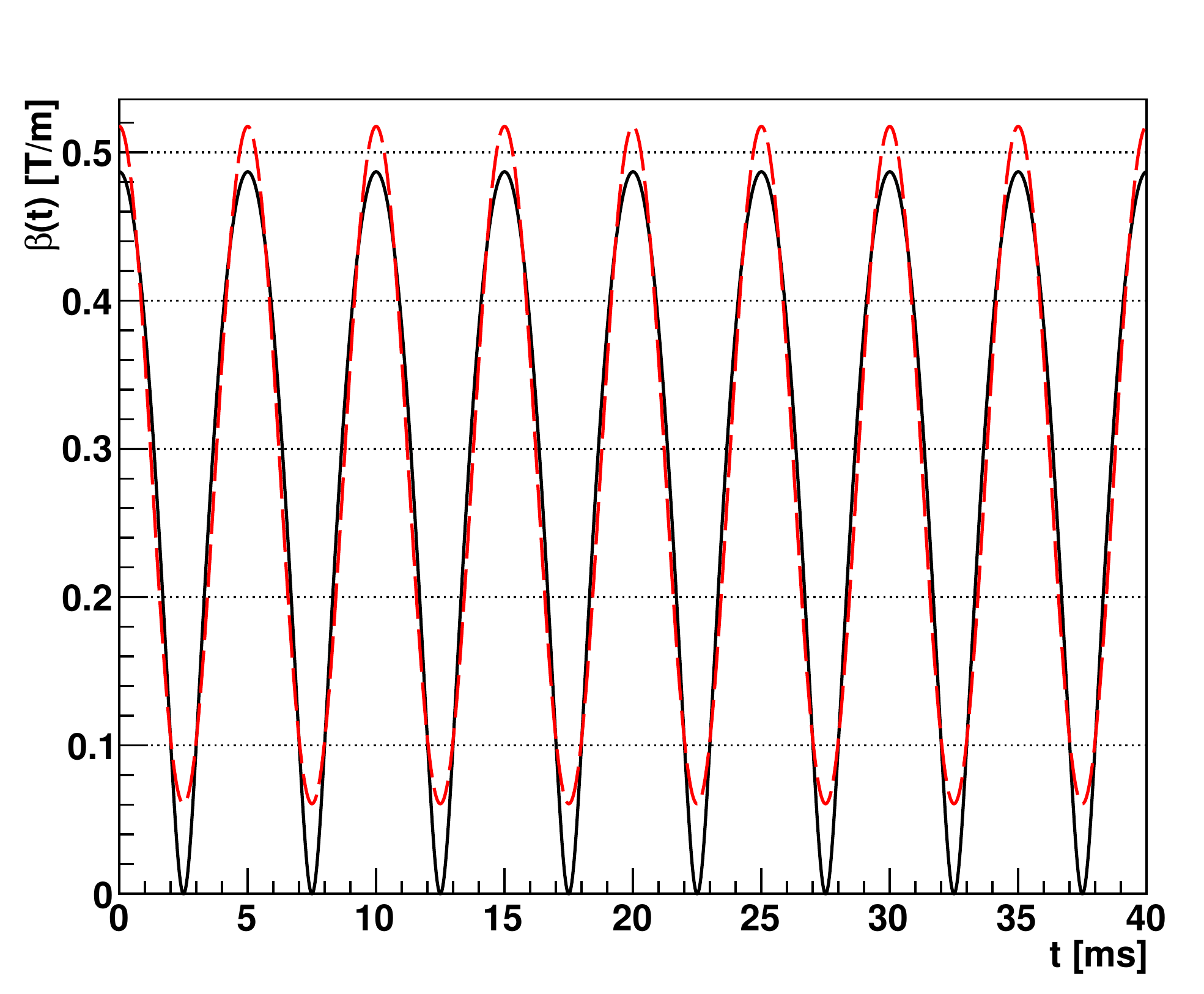}
\caption{
Time dependent gradient $\beta(t)$ for a driving frequency of $f = 100$~Hz, corresponding to an excitation frequency of $200$~Hz. 
Solid black line: eq. (\ref{grad2}), dashed red line: first order Fourier expansion eq. (\ref{Fourier}). 
}
\label{gradient_fourier}
\end{figure}

We have now defined a magnetic configuration for the AC mode with the wire array 
(oscillating currents at variable frequency $f$ and fixed amplitude $I_1 = I_4 = 1.4$~A, $I_2 = I_3 = 3.5$~A) 
and the external field ($B_{0, y} = 0.3$~mT) that 
(i) holds the neutron spin (ii) generate an oscillating gradient with the required amplitude to induce resonant transitions between quantum states. 
Note that the time dependent gradient $\beta(t)$ seen by the neutrons is not perfectly harmonic. 
Following eq. (\ref{grad}) the expression of the time dependent gradient is
\begin{equation}
\label{grad2}
\beta(t) = \partial_z |\vec{B}| = \hat{\beta} \ \frac{B_1 \cos^2(2 \pi f t + \phi) }{\sqrt{B_1^2 \cos^2(2 \pi f t + \phi) + B_{0, y}^2}}, 
\end{equation}
where $f$ is the driving frequency of the current in the wire array, $B_1 = 0.8$~mT and $\hat{\beta} = 0.52$~T/m.  
We plot $\beta(t)$ in fig. \ref{gradient_fourier}, where it is apparent that the excitation frequency (the frequency of the $\beta(t)$ excitation) is twice the driving frequency $f$ (the frequency of the oscillating currents in the wire array).

\begin{figure}
\includegraphics[width=0.72\linewidth,angle=90]{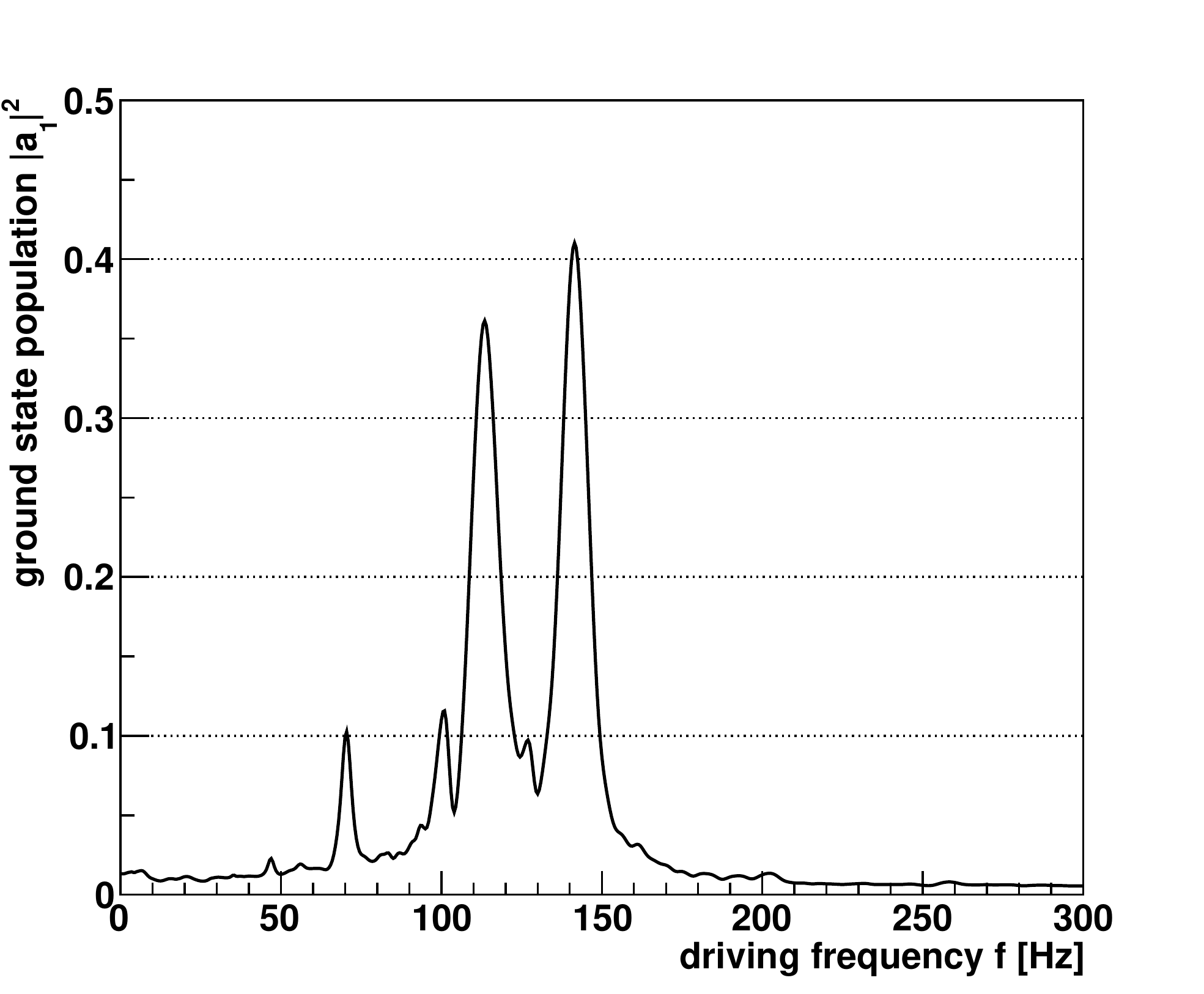}
\caption{
Numerical solution of the Schr\"odinger equation (\ref{schrodinger}). The probability of the $2 \rightarrow 1$ transition is plotted as a function of the driving frequency. 
}
\label{resonance}
\end{figure}

We will now simulate the transition probabilities that could be observed in the GRANIT flow through setup as described in section 2, assuming the benchmark magnetic configuration resulting in the excitation (\ref{grad2}). 
The problem consists in calculating the time evolution of a neutron quantum state 
\begin{equation}
| \psi(t) \rangle = \sum_n a_n(t) \ |n \rangle, 
\end{equation}
which is the solution of the time dependent Schr\"odinger equation
\begin{equation}
\label{schrodinger}
i \frac{d a_n}{dt} = E_n/\hbar \ a_n + \sum_m \frac{s}{2} \gamma \beta(t) \langle n|\hat{z}|m \rangle a_m. 
\end{equation}
Here we assume that the adiabaticity condition is fulfilled. 
The initial condition of the state is chosen to be $| \psi(0) \rangle = | 2 \rangle$ immediately after the preparation step. 
Then we solved eq. (\ref{schrodinger}) with a Runge-Kutta algorithm for a given set of parameters $f, v, s, \phi$, where the sum is restricted to the first four quantum states. 
For a given horizontal velocity $v$, the probability $|a_1(L/v)|^2$ for the neutron to be detected in the ground state at the exit of the magnetic excitation is calculated. 
The result is then averaged over the excitation phase $\phi$, the spin state $s = \pm 1$ and the velocity spectrum. 
We plot in fig. \ref{resonance} the transition probability as a function of the driving frequency $f$. 

According to this numerical calculation, we expect to see two resonances in the transmitted UCN flux associated with the $2 \rightarrow 1$ transition, with frequencies
\begin{equation}
f^+ = 141.5 \ {\rm Hz} \quad {\rm and} \quad f^- = 113.5 \ {\rm Hz}, 
\end{equation}
where $f^+$  and $f^-$ are the maxima of the resonance curve corresponding to spin ``up'' neutrons ($s=1$) and spin ``down'' neutrons ($s=-1$) respectively. 

The splitting of the resonances depending on the spin state could be interpreted as a ``Stern-Gerlach'' split due to a constant magnetic field gradient. 
To see this, it is useful to perform the Fourier expansion of the gradient excitation $\beta(t)$ given by (\ref{grad2})
\begin{equation}
\label{Fourier}
\beta(t) = \beta_0 + \beta_1 \cos(4 \pi f t + 2 \phi) + \cdots
\end{equation}
where $\beta_0 = 0.289$~T/m and $\beta_1 = 0.228$~T/m. 
The constant term $\beta_0$ of the excitation should be thought of as an spin-dependent effective modification of $g$. 
This is done by identification of the total constant vertical force (gravity plus constant gradient) to an effective gravitational force: 
\begin{equation}
m g^{\pm} = mg \pm \mu \beta_0. 
\end{equation}
Thus we expect a spin-dependent resonance frequency of the $n \rightarrow m$ transition given by
\begin{equation}
\label{SGshift}
f_{nm}^{\pm} = \frac{(m g^{\pm})^{2/3}}{2 \pi (2 m \hbar)^{1/3}} \ (\epsilon_n - \epsilon_m) = f_{n m} \ \left( 1 \pm \frac{\mu \beta_0}{mg} \right)^{2/3}
\end{equation}
where $f_{n m }$ is the unperturbed transition frequency. 
The splitting of the resonances obtained by the numerical resolution of the full Schr\"odinger equation (\ref{schrodinger}) is in quantitative agreement with eq. (\ref{SGshift}). 

In addition, eq. (\ref{SGshift}) motivates a combination of the two resonant frequencies $f^+$ and $f^-$ to extract the unperturbed transition frequency $f_{21}$, namely
\begin{equation}
f_{12} = \left( \frac{(2 f^+)^{3/2} + (2 f^-)^{3/2})}{2} \right)^{2/3}. 
\end{equation}
Remember that $f^+$ and $f^-$ refer to driving frequencies, that correspond to excitation frequencies of $2 f^+$ and $2 f^-$. 
When applied to the maxima of the curve shown in fig. \ref{resonance}, one extract $f_{21} = 255.8 \ {\rm Hz}$ which differs from the true resonance frequency $f_{21, {\rm true}} = 253.8 \ {\rm Hz}$ given by eq. (\ref{trueFrequencies}) by $2$~Hz. 
In fact there are several features of the full problem (given by eq. (\ref{schrodinger})), that the simple estimate (\ref{SGshift}) does not catch. 
The simplified formula (\ref{SGshift}) can be obtained by assuming a two level system (states $|1 \rangle$ and $|2 \rangle$) excited by an harmonic force given by (\ref{Fourier}) that couple states $|1 \rangle$ and $|2 \rangle$ 
and neglecting the self couplings of the type $\langle 1 | \hat{z} |1 \rangle$ and $\langle 2 | \hat{z} |2 \rangle$. 
The full calculation (\ref{schrodinger}) takes into account the non-harmonic excitation given by (\ref{grad2}) that couples all states $|1 \rangle$, $|2 \rangle$, $|3 \rangle$ and $|4 \rangle$ including self couplings. 
All these complications are potential sources of frequency shifts of the resonance line. 
We then conclude that these shifts are below the percent level.

\section{Conclusion}
In the GRANIT flow through arrangement, two possible modes to induce resonant transitions between the quantum states could be used. 
In the DC mode, where the magnetic field gradient oscillates spatially along the $x$ direction, the excitation frequency is controlled by the horizontal neutron velocity. 
In the AC mode, where the gradient is homogeneous in space and oscillates in time, the excitation frequency is selected directly by the frequency of the current driving the magnetic excitation. 
We have shown that the condition of adiabaticity of spin transport can also be fulfilled in the AC mode using a moderate horizontal magnetic field normal to the neutron propagation axis. 
Finally a calculation of the expected resonance line for the $2 \rightarrow 1$ transition indicates that a measurement of the transition frequency at a precision better than a percent is possible. 
A detailed comparison describing the relative merits of the two methods with the associated systematic effects is left for a future work.

\appendix
\section{Magnetic field of a square wire}

Here we provide formulas for the magnetic field generated by an infinitely long square wire with current $I$ flowing uniformly in the wire in the $y$ direction. 
The magnetic field lies in the $(x,z)$ plane. 
We assume that $(x=0, y=0)$ corresponds to the center of the wire. 
The length of the square is denoted by $c$ ($c = 1$~mm for the purpose of this article). 
Formulas are valid outside the wire. 
We define the following quantities
\begin{eqnarray*}
x_m & = x - c/2 ,  \quad  z_m & = z - c/2,  \\
x_p & = x + c/2 ,  \quad  z_p & = z + c/2,  
\end{eqnarray*}
\begin{eqnarray*}
L_1 & = & \ln \left( \frac{(x_m^2 + z_m^2) (x_p^2 + z_p^2) }{(x_p^2 + z_m^2) (x_m^2 + z_p^2)} \right), \\
L_2 & = & \ln \left( \frac{(x_m^2 + z_m^2) (x_p^2 + z_m^2) }{(x_m^2 + z_p^2) (x_p^2 + z_p^2)} \right), \\
L_3 & = & \ln \left( \frac{(x_m^2 + z_m^2) (x_m^2 + z_p^2) }{(x_p^2 + z_m^2) (x_p^2 + z_p^2)} \right).
\end{eqnarray*}
\begin{eqnarray*}
A_{mm} & = \arctan(x_m/z_m), \ A_{pm} & = \arctan(x_p/z_m), \\
A_{mp} & = \arctan(x_m/z_p), \ A_{pp} & = \arctan(x_p/z_p). 
\end{eqnarray*}
By integrating the Biot-Savart law over the volume of the wire we find the following expressions for the field components
\begin{eqnarray}
B_x(x,z) & =      &  - \frac{\mu_0 I}{ 4 \pi c^2} \ \big[ x L_1 - \frac{c}{2} L_2 +  \\ 
\nonumber
	 &  &  2 z_m \left( A_{mm} - A_{pm} \right) + 2 z_p \left( A_{pp} - A_{mp} \right) \big],  \\
B_z(x,z) & =      &  \frac{\mu_0 I}{ 4 \pi c^2} \ \big[ z L_1 - \frac{c}{2} L_3 +  \\ 
\nonumber
	 &  &  2 x_m \left( A_{mp} - A_{mm} \right) + 2 x_p \left( A_{pm} - A_{pp} \right) \big].
\end{eqnarray}
By taking the derivative we find the following expressions for the gradients
\begin{eqnarray}
\partial_z B_x & = & \frac{\mu_0 I}{2 \pi c^2} \ \big[ A_{mp} - A_{mm} + A_{pm} - A_{pp} \big], \\
\partial_z B_z & = & \frac{\mu_0 I}{4 \pi c^2} \, L_1.
\end{eqnarray}

\bibliographystyle{unsrt}

\end{document}